\documentclass[lettersize,journal]{IEEEtran}
\usepackage{cite}
\usepackage{amsmath,amssymb,amsfonts}
\usepackage{graphicx}
\usepackage{textcomp}
\usepackage{booktabs}
\usepackage{multirow}
\usepackage{cite}
\makeatletter
\let\NAT@parse\undefined
\makeatother
\usepackage{hyperref}
\hypersetup{hypertex=true, colorlinks=true, linkcolor=blue,	anchorcolor=blue, citecolor=blue}
\usepackage[capitalize]{cleveref}

\usepackage{array}
\usepackage[caption=false,font=normalsize,labelfont=sf,textfont=sf]{subfig}
\usepackage{stfloats}
\usepackage{url}
\usepackage{verbatim}
\begin{document}

\title{Dc-EEMF: Pushing depth-of-field limit of photoacoustic microscopy via decision-level constrained learning}

\author{Wangting Zhou, Jiangshan He,Tong Cai, Lin Wang, Zhen Yuan, Xunbin Wei, Xueli Chen
	\thanks{This work was supported in part by the National Natural Science Foundation of China under Grant 62105255, Grant 62275210; in part by the Key Research and Development Program of Shaanxi under Grant 2023-YBSF-293; in part by the Xidian University Specially Funded Project for Interdisciplinary Exploration under Grant TZJH2024043; in part by the Fundamental Research Funds for the Central Universities under Grant 20101257147; in part by the National Key R\&D Program of China under Grant 2022YFB3203800; and in part by National Young Talent Program, and Shaanxi Young Top-notch Talent Program. (Wangting Zhou and Jiangshan He contributed equally to this work.) (Corresponding authors: Xunbin Wei and Xueli Chen.)}
	\thanks{Wangting Zhou, Jiangshan He, and Xueli Chen  are with the Center for Biomedical-photonics and Molecular Imaging, Xi’an Key Laboratory of Intelligent Sensing and Regulation of trans-Scale Life Information, School of Life Science and Technology, Xidian University, Xi'an, Shaanxi 710126, China. (e-mail: wtzhou@xidian.edu.cn; hejs9603@163.com; xlchen@xidian.edu.cn) }
	\thanks{Lin Wang is with the School of Computer Science and Engineering, Xi’an University of Technology, Xi’an, Shaanxi 710048, China. (wanglin004@xaut.edu.cn)}
	\thanks{Zhen Yuan is with the Faculty of Health Sciences, University of Macau, Macau, 999078, China. (zhenyuan@um.edu.mo)}
	\thanks{Xunbin Wei is with the Laboratory of Carcinogenesis and Translational Research (Ministry of Education/Beijing), Peking University Cancer Hospital \& Institute, Beijing, China, also with Biomedical Engineering Department, Peking University, Beijing 100081, China. (xwei@bjmu.edu.cn)}}

\markboth{XXXXXXXXXXXXXXX}%
{Shell \MakeLowercase{\textit{et al.}}: A Sample Article Using IEEEtran.cls for IEEE Journals}


\maketitle

\begin{abstract}
Photoacoustic microscopy holds the potential to measure biomarkers' structural and functional status without labels, which significantly aids in comprehending pathophysiological conditions in biomedical research. However, conventional optical-resolution photoacoustic microscopy (OR-PAM) is hindered by a limited depth-of-field (DoF) due to the narrow depth range focused on a Gaussian beam. Consequently, it fails to resolve sufficient details in the depth direction. Herein, we propose a decision-level constrained end-to-end multi-focus image fusion (Dc-EEMF) to push DoF limit of PAM. The DC-EEMF method is a lightweight siamese network that incorporates an artifact-resistant channel-wise spatial frequency as its feature fusion rule. The meticulously crafted U-Net-based perceptual loss function for decision-level focus properties in end-to-end fusion seamlessly integrates the complementary advantages of spatial domain and transform domain methods within Dc-EEMF. This approach can be trained end-to-end without necessitating post-processing procedures. Experimental results and numerical analyses collectively demonstrate our method's robust performance, achieving an impressive fusion result for PAM images without a substantial sacrifice in lateral resolution. The utilization of Dc-EEMF-powered PAM has the potential to serve as a practical tool in preclinical and clinical studies requiring extended DoF for various applications.
\end{abstract}


\section{Introduction}
\par\IEEEPARstart{P}{hotoacoustic} microscopy (PAM) is a hybrid imaging technique that employs optical excitation and ultrasound detection. It combines the high-contrast and high-resolution advantages of optics with the deep-penetration capabilities of acoustics, making it appealing for various clinical and pre-clinical applications \cite{1_da_mesquita_functional_2018,2_louveau_cns_2018,3_wang_multiscale_2009}. In PAM, a pulsed laser serves as the excitation source to target the absorber. The photo-thermal effect induces a temperature rise, generating an acoustic pressure rise via thermo-elastic expansion, directly proportional to the original optical absorption of biological tissue \cite{4_xu_erratum_2007}. Subsequently, the acoustic pressure wave propagates through the tissue and is detected by an ultrasonic transducer, forming an image based on optical absorption.

\par\textcolor{black}{In optical-resolution PAM (OR-PAM), microscopic resolution is achieved by focusing a Gaussian beam into a micron-level spot. However, due to light diffraction, the focused Gaussian beam has a limited depth range in focus, creating a tradeoff between lateral resolution and the depth of field (DoF). This tradeoff affects the characterization of pathophysiological biomarkers by limiting the level of detail in the depth direction \cite{5_gong_photoacoustic_2022}. To address the out-of-focus issue, several methods have been developed to extend the DoF. Hardware based approaches to improve DoF include utilizing the nondiffracting property of a Bessel beam or a needle-shaped beam, which offer a larger DoF compared to a Gaussian beam \cite{6_hu_extended_2019,7_zhou_deep_2022, 54_Cao_2023}. However, the artifacts introduced by the side lobes of the Bessel beam require suppression using non-linear methods, and the laser energy density of the main lobe is typically weak. The modulation efficiency of the needle-shaped beam is also relatively low. Electrically tunable lenses or liquid lenses can also enhance the DoF by shifting the light focus back and forth axially \cite{8_lee_high-speed_2017,9_liu_continuous_2022}. However, their response time often fails to keep up with lasers operating at MHz repetition rates due to physical limitations, inertia, mechanical constraints, thermal effects, and optimization trade-offs \cite{7_zhou_deep_2022}. Additionally, applying electrically adjustable lenses or liquid lenses faces challenges in control, calibration, alignment, and integration, impacting the practicality of OR-PAM and complicating the system to some extent \cite{10_zuo_high-speed_2013}. Similarly, other hardware methods may offer advantages in expanding the DoF \cite{55_Li_2023, 53_Liang_2023}, this enhancement inevitably introduces greater system complexity and elevated costs.}

\par\textcolor{black}{Multi-focus image fusion technique emerges as an effective strategy for overcoming the limitations imposed by the limited DoF inherent in optical imaging systems. By fusing multiple partially focused image stacks into a single, all-in-focus image, this technique enables the creation of images with enhanced detail and clarity. In our previous work, we proposed a computed multi-focus image fusion method to extend the DoF in intravital OR-PAM and demonstrated its robustness in quantifying vasculatures \cite{11_zhou_multi-focus_2022}. Nevertheless, despite advancements in scanning techniques, challenges such as inaccuracies in device calibration and the inevitable effects of animal respiration and cardiac activity persist. These factors can lead to incomplete registration of source images, thereby compromising the overall quality of the fused image. Addressing these issues is crucial to enhance the accuracy and reliability of diagnostic imaging, ensuring more discerning and meticulous insights into biological processes and pathology. Recently, deep learning-enabled multi-focus image fusion has garnered attention in natural images and optical imaging systems for extended DoF tasks. 
These methods are categorized into decision-map-based methods \cite{60_DRPL_2020, 64_DRL_FPD_2022} and end-to-end-based methods \cite{61_IFCNN_2020, 62_U2Fusion_2022, 63_MFFGAN_2021, 65_SWinFusion_2022, 66_MUFusion_2023,67_DBMFIF_2024}. The former, akin to spatial-domain-based methods in traditional algorithms \cite{57_MWGF_2014,58_GFDF_2019, 59_MWGIF_2022}, tends to preserve the original image's focus information to a great extent, contributing to the sharpness of the fused image.
However, this copy-and-paste approach can introduce undesirable artifacts at focus boundaries, and the variation in sharpness among source images can pose significant challenges for accurate boundary detection, as encountered in our previous work. The latter methods share a framework similar to transform-domain-based methods in traditional algorithms \cite{56_NSCT_LP_2015}, using a convolutional network to achieve an end-to-end mapping from source images to the fused image. While these methods excel in retaining information from the source images, there may be some intensity degradation in the sharpness of the fused images. Several attempts have been made to combine the strengths of these approaches. For instance, DRL-FPD \cite{64_DRL_FPD_2022} integrates complementary advantages in decision-map-based methods and end-to-end-based methods, and the focus property detection accuracy is significantly improved. However, this method still faces challenges in preserving information at the focus-defocus boundaries due to its reliance on the decision-map-based MFIF pipline.}

\par\textcolor{black}{In this article, we present a decision-level constrained end-to-end multi-focus image fusion (Dc-EEMF) to push DoF limit for OR-PAM. We demonstrate its robust performance in characterizing vasculatures using a deep learning-powered multi-focus image fusion approach. Inspired by the complementary advantages of transform domain-based methods and spatial domain-based methods, we designed a deep convolutional network to extract essential information while discarding irrelevant elements. Initially, we utilized simulation data to validate the effectiveness of our proposed methodology. Subsequently, phantom and in vitro liver tissue experiments were conducted to validate the feasibility of the Dc-EEMF model in handling real PAM data. The goal was to retain the original focus information from rapid raster scanning of different focus-planes as well as prevent visual artifacts that frequently occur at the boundaries between focused and de-focused areas. This led to superior qualitative and quantitative performance when compared to state-of-the-art computed methods. Lastly, cortex-wide imaging was performed on in vivo mouse brains to demonstrate the potential of Dc-EEMF-powered extended-DoF PAM in comprehensive vasculature quantification within uneven brain topological structures. The primary contributions of this work are outlined as follows:}
	
\par\textcolor{black}{ 1. A decision-level constrained end-to-end learned multi-focus image fusion (Dc-EEMF) method is proposed to push the DoF limit for OR-PAM. To the best of our knowledge, it is unprecedented in the existing deep learning-based works in this field.}

\par\textcolor{black}{ 2. The decision-level focus property perceptual loss function for the end-to-end fusion is intricately designed to integrate the complementary advantages of spatial domain methods and transform domain methods in Dc-EEMF. A channel-wise spatial frequency employed as the feature fusion rule is proposed to mitigate the impact of artifacts during the Dc-EEMF processing.}

\par\textcolor{black}{ 3. Experimental results demonstrate that the proposed method can achieve competitive performance in terms of both visual quality and objective evaluation when compared with state-of-the-art methods. This Dc-EEMF-powered PAM contributes to imaging applications of various topological structures of surfaces.}

\section{Methods}

\begin{figure*}[!t]
	\centering
	\includegraphics[width=180mm]{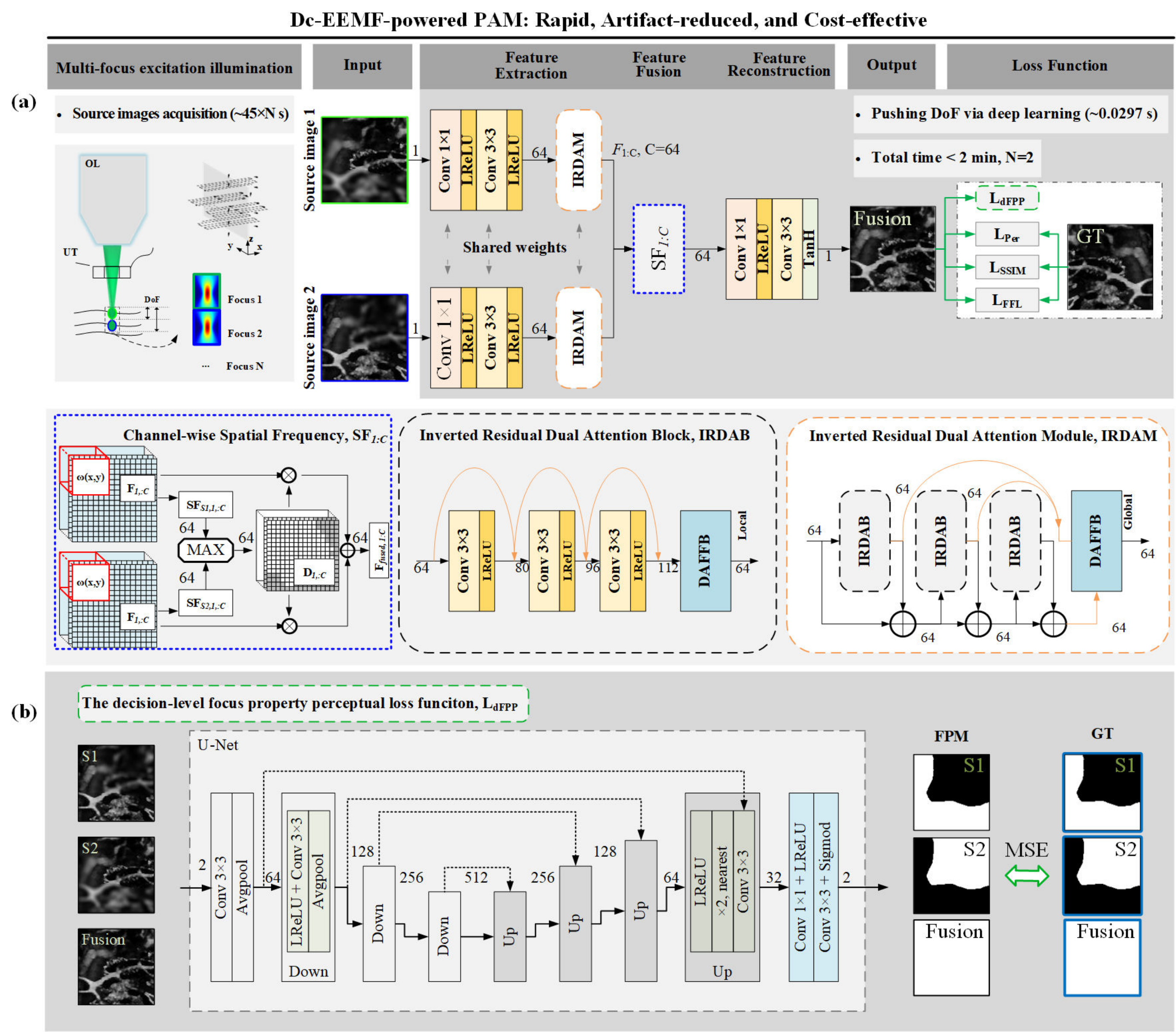}
	\caption{The proposed Dc-EEMF framework based on a convolutional neural network. (a) Architecture of the Dc-EEMF learned network consisting of feature extraction, feature fusion, and feature reconstruction. Further details about the submodules of Dc-EEMF are provided below the framework. $L_{dFPP}$, $L_{Per}$, $L_{SSIM}$,  $L_{FFL}$ denote decision-level focus property perceptual loss, perceptual loss function, structural similarity index measure loss, and focal frequency loss, respectively. (b) The U-Net-based dFPP loss function give a decision level constrict for Dc-EEMF network. FPM, focus property map. MSE, mean squared error. LReLU, leaky rectified linear unit with 0.2 negative slope. \textcolor{black}{DAFFB, the multiplicative dual-attention (channel and spatial) feature fusion block.} GT, ground truth. OL, objective lens. S1, S2, source image 1, and source image 2. UT, ultrasound transducer.}
	\label{figure1}
	
\end{figure*}

\subsection{Concept and architecture of the Dc-EEMF-powerd PAM}
\par\textcolor{black}{Drawing inspiration from both spatial domain-based methods and transform domain-based methods, namely, decision-map-based methods and end-to-end-based methods in deep learning-powered algorithms, this work focuses on integrating the complementary advantages of these two categories of methods. It introduces a Dc-EEMF method to push DoF limit for photoacoustic microscopy. Specifically, we employ U-Net-based loss functions to learn the properties of focus area distribution. Subsequently, we fuse source images with different focuses in an end-to-end manner to attain more precise fusion details. This approach allows for the coordination and combination of the complementary advantages of decision-map-based methods and end-to-end methods. Moreover, a channel-wise spatial frequency-based feature fusion rule is designed to further diminish the impact of artifacts during the Dc-EEMF processing. Figure \ref{figure1} illustrates the schematic diagram of the proposed Dc-EEMF-powered PAM.}

\subsubsection{Network architecture}
\par\textcolor{black}{The proposed network comprises three components: feature extraction, feature fusion, and feature reconstruction. In all convolution stages, the feature maps' size aligns with the size of source images, and the channel dimension of feature maps is indicated in the network structure. Feature extraction is a crucial procedure in multi-focus image fusion methods, we begin by employing two convolution layers with LReLU activation functions to extract the shallow features from the source images. Inspired by \cite{34_howard_searching_2019,35_woo_cbam_2018}, we design the inverted residual dual-attention module (IRDAM) to dynamically learn extensive and informative image features in high dimensions at both local and global hierarchical levels before projecting them onto a low-dimensional representation (Fig. \ref{figure1}(a)). The input for the feature extraction comprises paired source images, while the output yields paired \textit{C}-dimensional vectors of the feature map ${F}_{1:C}$.}

\par\textcolor{black}{Feature fusion is essential to retain sharp texture details while discarding blurry textures for the multi-focus image fusion task. To effectively suppress slightly misaligned artifacts and extract crucial focus region information, we utilized the channel-wise spatial frequency (${SF}_{1:C}$) as a feature fusion rule to gauge the focus activity levels of the feature map ${F}_{1:C}$ derived from the IRDAM. Additionally, the window-based accumulation strategy is applied to mitigate noise interference. The channel-wise spatial frequency can be formulated as follows:}

\color{black}
\begin{equation}
	{SF}_{1:C}\left( {x,y} \right) = {\sum\limits_{{({i,j})} \in \omega{({x,y})}}\sqrt{{RF}_{1:C}^{2}\left( {i,j} \right) + {CF}_{1:C}^{2}\left( {i,j} \right)}}
\end{equation}

\par\textcolor{black}{ Where the row vector frequencies ${RF}_{1:C}\left( {i,j} \right) = \left| {F_{1:C}\left( {x,~y} \right) - F_{1:C}\left( {x,~y - 1} \right)} \right|$, the column vector frequencies ${CF}_{1:C}\left( {i,j} \right) = \left| {F_{1:C}\left( {x,~y} \right) - F_{1:C}\left( {x - 1,y} \right)} \right|$. The window-based strategy exhibits significant impedance with the artifacts introduced by the boundary regions, and $\omega\left( {x,y} \right)$ is a small window centered at pixel $(x,y)$ with a size set to $11\times11$.}
\par\textcolor{black}{The feature decision tensor ${D}_{1:C}(x,y)$ is obtained by comparing the spatial frequencies of the source images ${SF}_{S1,1:C}$ and ${SF}_{S2,1:C}$ as follows:}

\begin{equation}
	D_{1:C}\left( {x,y} \right) = \left\{ \begin{matrix}
		1 & {if~SF_{S1,1:C}\left( {x,~y} \right) \geq SF_{S2,1:C}\left( {x,~y} \right)} \\
		0 & {otherwise}
	\end{matrix} \right.
\end{equation}

\par\textcolor{black}{Hence, the fused feature map $F_{fused,1:C}$ can be determined as:} 

\begin{equation}
	F_{fused,1:C} = F_{S1,1:C} \cdot D_{1:C} + F_{S2,1:C} \cdot \left( {1 - D_{1:C}} \right).
\end{equation}

\par\textcolor{black}{Finally, a concise two-layer convolutional neural network is designed to reconstruct fusion features, which reduces the number of channels to the same as the input.}

\subsubsection{The loss functions}
\par\textcolor{black}{To effectively optimize the Dc-EEMF network, a strategy that combines multiple loss functions is employed, which preserves and reconstructs focus region information from the perspectives of decision-level focus property distribution, texture perceptual quality, spatial features, and high-frequency information in spectral space. Here, we utilize a combination of decision-level focus property perceptual (dFPP) loss  as shown in Fig. \ref{figure1}(b), perceptual loss, structural similarity index measure (SSIM) loss, and focal frequency loss. This amalgamation results in a total loss that yields the best performance in our network.}

\begin{equation}
	L_{total} = L_{dFPP}+{\alpha_{1}L}_{per}+{\alpha_{2}L}_{SSIM}+{\alpha_{3}L}_{FFL}
\end{equation}

\par\textcolor{black}{ Where $\alpha_*$ is used to adjust these loss terms to the same level of importance, here, we set $\alpha_1=0.2$, $\alpha_2=1$, $\alpha_3=8$.}
\par\textcolor{black}{Initially, the U-Net-based decision-level focus property perceptual loss function, which is trained on our simulation datasets to detect focus region information. The $L_{dFPP}$ is defined as follows:}

\begin{equation}
\text{ $L_{dFPP}$} = \sum_{S_{i} \in \{ S1, S2 \}} \text{MSE} \left[ U_{\text{pred}} \left( I_{\text{Fused}}, I_{S_i} \right), FPM_{\left( \text{GT}, S_{i} \right)} \right]
\end{equation}

\par\textcolor{black}{ where $FPM_{(GT, S_i)}$ denote a focus property map collections of the ground truth image and source image S1, S2. These maps contain accurate information regarding the focus and defocus regions.}

\par\textcolor{black}{ Next, to further enhance the images' texture similarity between the model output and the ground truth, a perceptual loss $L_{per}$ based on the pre-trained VGG19 network is used, which can keep the image details more accurate \cite{37_johnson_perceptual_2016}.} 

\begin{equation}
	L_{per} = \left\| {\phi_{j}\left( I_{fused} \right) - \phi_{j}\left( I_{GT} \right)} \right\|_{2}
\end{equation}

\par\textcolor{black}{ Where $\phi$ is the feature map of the j-th layers of pre-trained VGG19 on the fused image $I_{fused}$ or the ground-truth $I_{GT}$, and $j\in[2^{nd},7^{th},12^{th},21^{th},30^{th} ]$. The single-channel fused image is copied to form RGB channels to match the VGG network input channel.}
	
\par\textcolor{black}{The third component is the SSIM loss function $L_{SSIM}$, which is considered one of the most popular and efficient metrics for adjusting spatial details distortion between the fused image and ground truth based on similarities in luminance, contrast, and structural information. It can be formulated as follows:}

\begin{equation}
	L_{SSIM} = 1 - SSIM_{fused,GT}
\end{equation}

\par\textcolor{black}{The final component is the focal frequency loss $L_{FFL}$, which provides the network with additional frequency information and enhances the quality of image reconstruction. The focal frequency loss is defined as follows \cite{38_jiang_focal_2021}:}

\begin{equation}
	L_{FFL} = \frac{1}{WH}{\sum\limits_{u = 1}^{W}{{\sum\limits_{v = 1}^{H}{\omega\left( {u,v} \right)}}\left| {F_{fused}\left( {u,v} \right) - F_{GT}\left( {u,v} \right)} \right|^{2}}}
\end{equation}

\par\textcolor{black}{ Where $F_{fused}$ and $F_{GT}$ are the two-dimensional discrete Fourier transforms of the fused image and ground truth image, respectively. $\omega(u,v)$ represents the spectrum weight matrix, which is dynamically determined by a non-uniform distribution based on the normalized current loss of each frequency during training. And $\omega(u,v)$ is defined as the normalized euclidean distance between $F_{fused}$ and $F_{GT}$.}

\subsection{Experimental details}

\subsubsection{Datasets for model training and initial test}
\par\textcolor{black}{We adopt a simulated data generation strategy, which not only addresses the problem of limited training data, but also provides ground truth for training. The photoacoustic microscopy datasets as mentioned in \cite{44_chen_simultaneous_2019} were used, and the multi-focus image dataset is generated from PAM images and randomly generated FPM, simultaneously, according to Gaussian blurring \cite{26_zhang_ifcnn_2020}. The original data of PAM (2000$\times$1200, 1600$\times$600, 1800$\times$1390 and so on) is cropped into size of 128$\times$128 pixels. To enhance the performance of the model, the data augmentation techniques are achieved by performing the random horizontal or vertical rotation, as well as the randomly cropping the image with size of 64$\times$64 pixels.}

\subsubsection{Dc-EEMF-powered PAM imaging setups}
\par\textcolor{black}{The Dc-EEMF-powered PAM utilizes a nanosecond pulsed 532 nm laser (POPLAR-532-10B, Huaray; repetition rate 1 KHz). The ultrasonically excited waves are detected by the ring transducer (UT, focal length: 8.4 mm; center frequency: 50 MHz; 6-dB bandwidth: 90\%), amplified by two 30-dB amplifiers (ZFL-500LN+, Mini-Circuits), filtered through a 48-MHz low-pass filter (SLP-50+, Mini-Circuits), and acquired by a high-speed data acquisition board (DAQ, PCIE-1425, Yixing Technology) at a 250 MS/s sampling rate. The PAM multi-focus source images are obtained from two consecutively rapid raster scans covering different focus area.}

\begin{figure*}[!t]
	\centering
	\includegraphics[width=160mm]{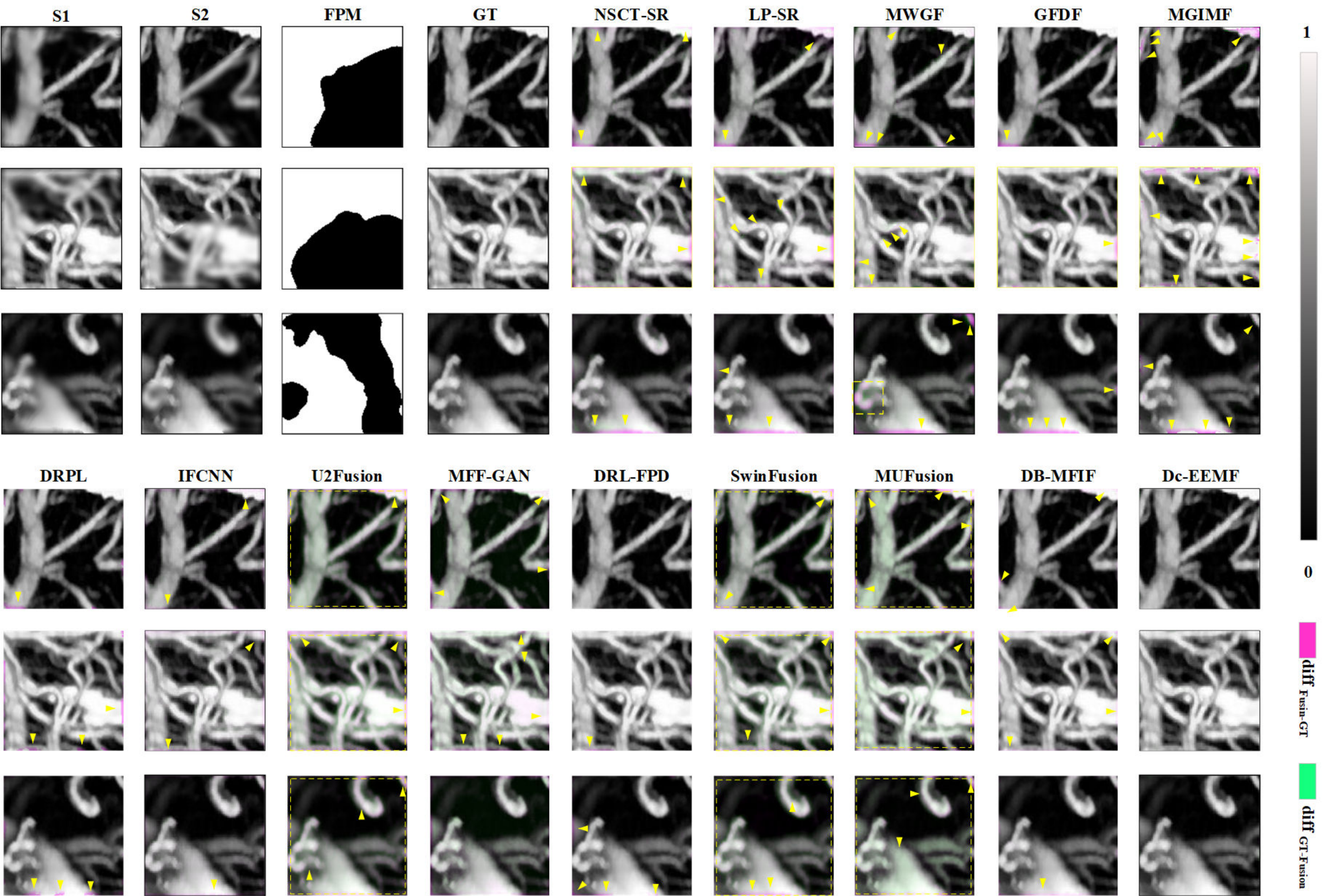}
	\caption{\textcolor{black}{Simulation study with the proposed Dc-EEMF from two axial layers. The complete-cleared image (ground truth), the focus property map, two source images, and a comparison of fusion results among different methods which including 2 transform-domain methods, NSCTSR, LP-SR, and 3 spatial-domain methods, MWGF, GFDF, MGIMF, and 8 deep learning methods, DRPL, IFCNN, U2Fusion, MFF-GAN, DRL-FPD, SwinFusion, MUFusion, DB-MFIF. Fusion artifact are indicated by yellow arrows and yellow dash boxes. Diff. denotes the difference image between fusion results (pink areas) and GT (green areas).}}
	\label{figure3}
\end{figure*}

\subsubsection{Evaluation metrics}
\par\textcolor{black}{To evaluate the performance of the fusion methods comprehensively, we adopt four often used metrics to evaluate the quantitative performance of the algorithms \cite{46_zhang_deep_2021, 47_liu_multi-focus_2020}: the structural similarity-based metric ($Q_E$), the human visual perception-based metric ($Q_{CV}$), the phase congruency-based metric ($Q_P$), and standard deviation (SD). Note that, for each metric except for $Q_{CV}$, a larger value indicates a better fusion quality. However, smaller absolute value of $Q_{CV}$ demonstrates the better fused result. Furthermore, in order to comprehensively amalgamates several metrics $Q_{CV}$, $Q_E$, $Q_P$, and SD, the Borda count score  \cite{46_zhang_deep_2021} is employed. In all the metrics, the first to last place gets N to 1 point, respectively, and N is the total of ranking tasks. Each metric is treated equally, and a larger value indicates a better performance.}

\subsubsection{Training details}
\par\textcolor{black}{To ensure the reproducibility of our experiments, we detail the key components of our training environment. For the training process, we conducted all experiments using Pytorch 1.11.0, torchvision 0.12.0, and CUDA 12.2. The Adam solver is used with parameters $\beta_1=0.9$, $\beta_2=0.99$ and $\epsilon=10^{-8}$. The learning rate is set to $1e^{-3}$, and an exponential decay of rate 0.9 every 5 epochs is used both. The batch size is set to 16. The model is trained on the generated dataset for 120 epochs. The time consummation of a paired multi-focus images with 128$\times$128 size is about $2.97\times10^{-2}$ s.}

\subsubsection{Training details for U-Net-based decision-level focus property perceptual loss}
\par\textcolor{black}{ The structure of dual-input encoding-decoding network \cite{45_jo_investigating_2020} is shown in Fig. \ref{figure1}(b), which is used to identify the FPM, in other words, the focus information distributions of the paired input images. Upon completion of the training process, the U-Net model achieved a mean intersection over union (MIoU) score of 0.983, indicating a high level of segmentation accuracy. However, this performance still suggests the presence of minor misclassified regions within the segmented outputs. That’s the reason why the spatial domain-based methods always need the post-processing step, as well as the decision-map-based methods. In the training phase, the output are the focus property maps with same size as the input. And the U-Net training loss function is:}

\color{black}
\begin{equation}
Loss = \sum\limits_{i,j\in\{{S1},{S2},GT\}}{MSE\left[{{U_{pred}}\left( {i,j} \right),FPM_{(i,j)}} \right]}
\end{equation}


\par\textcolor{black}{ Where $FPM_{(i,j)}$ represent a focus property map collections of the ground truth image $I_{GT}$, source image S1 and S2, which contains ground truth information about the focus and the defocus region.}

\begin{figure*}[!t]
	\centering
	\includegraphics[width=150mm]{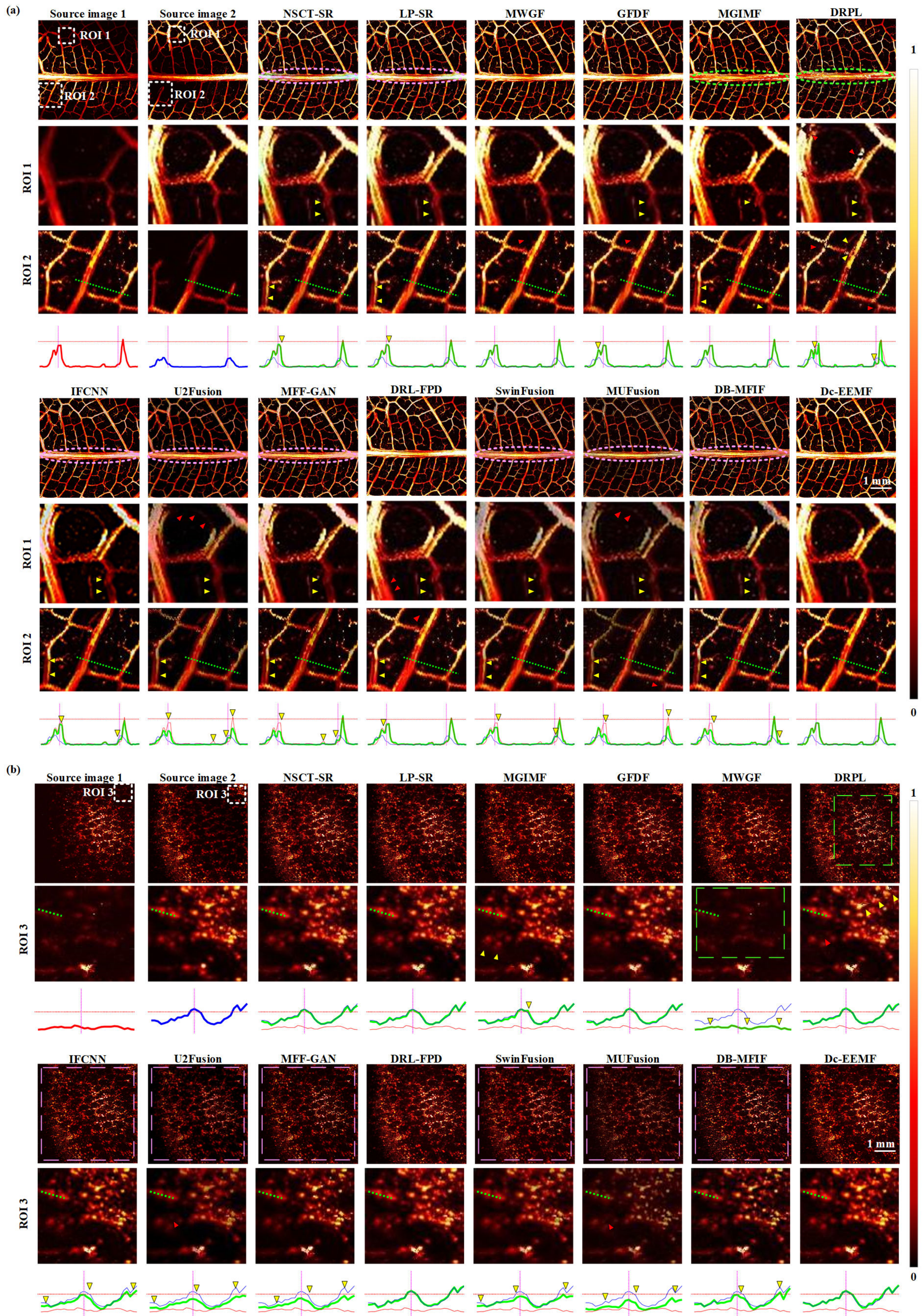}
	\caption{\textcolor{black}{Phantom and in vitro study with Dc-EEMF. (a) and (b) Source images (S1, S2) and the fusion results of leaf skeleton and liver tissue obtained by different methods which including 2 transform-domain methods, NSCTSR, LP-SR, and 3 spatial-domain methods, MWGF, GFDF, MGIMF, and 8 deep learning methods, DRPL, IFCNN, U2Fusion, MFF-GAN, DRL-FPD, SwinFusion, MUFusion, DB-MFIF. Pink circles/boxes represent image reconstruction errors. Green circles/boxes indicate low-quality fusion results caused by suboptimal focus regions in spatial-based methods. Yellow arrows highlight fusion artifacts. Red arrows denote information omission. In the line ROI analysis, red/blue lines correspond to the original source images, while other green lines represent ROIs generated by different fusion methods for comparative purposes. Source images of leaf skeleton are obtained at depths of 250 and 750 µm. Source images of liver tissue are obtained at depths of 250 and 500 µm.	
}} 
	\label{figure4}
\end{figure*}

\section{Results}

\subsection{Simulation study via Dc-EEMF-powered PAM}

\par\textcolor{black}{ To validate the effectiveness of the proposed Dc-EEMF method, an extended DoF study is conducted using the PAM multi-focus simulation dataset (100 groups). 
Here, we performed a comprehensive comparison of 13 methods, including 2 transform-domain methods, NSCT-SR(2015) \cite{56_NSCT_LP_2015}, LP-SR(2015)\cite{56_NSCT_LP_2015}, and 3 spatial-domain methods, MWGF(2014)\cite{57_MWGF_2014}, GFDF(2019)\cite{58_GFDF_2019}, MGIMF(2022)\cite{59_MWGIF_2022},and 8 deep learning methods, DRPL(2020)\cite{60_DRPL_2020}, IFCNN(2020)\cite{61_IFCNN_2020}, U2Fusion(2020)\cite{62_U2Fusion_2022}, MFF-GAN(2021)\cite{63_MFFGAN_2021}, DRL-FPD(2022)\cite{64_DRL_FPD_2022}, SwinFusion(2023)\cite{65_SWinFusion_2022}, MUFusion(2023)\cite{66_MUFusion_2023}, DB-MFIF(2024)\cite{67_DBMFIF_2024}.
Representative fusion results are presented in Fig. \ref{figure3}.
Evidently, upon subjective inspection, it's noticeable that the spatiail-domain-based and decision-map-based methods, such as MWGF, GFDF, MGIMF, and DRPL, exhibit irregular regions of mis-fusion artifacts and information omission, indicated by yellow arrows and box (Detailed areas need to be zoomed in to observe). And the transform-domain-based and end-to-end-based methods, such as NSCT-SR, LP-SR, DRPL, IFCNN, U2Fusion, MFF-GAN, DRL-FPD, SwinFusion, MUFusion, and DB-MFIF, exhibit areas of intensity change and mis-fusion artifacts. And our proposed Dc-EEMF performs the best. Quantitatively, the PSNR, SSIM, and MSE evaluation metrics are calculate between ground truth and fusion results among different methods. In Table \ref{table3}, the  Dc-EEMF method ranked the first place in all metrics, which demonstrates the superiority of the proposed method in terms of the fusion quality, the level of reconstruction error reduction, and the preservation of source images' focus spatial features for the fused images. Through the aforementioned fused results and analysis, the Dc-EEMF method is proven to deliver the best performance both qualitatively and quantitatively from the subjective visual perception and objective metrics.}

\begin{table*}[t]
	\setlength{\abovecaptionskip}{0cm} 
	\setlength{\belowcaptionskip}{-1cm}
	\caption{Statistical Metrics (Mean±SD) for Simulation Study}

	\label{table3}%
	\begin{center}
		\begin{tabular}{ccccc}
			\toprule[0.3mm]
			Metrics & {PSNR} & {MSE} & {SSIM} & {(Rank) Borda count}\\
			\midrule[0.3mm]
			{NSCT-SR(2015)} & (9) 0.9773 ± 0.0155 & (7) 34.5269 ± 37.8541 & (9) 34.9982 ± 4.4914 & (9) 19 \\[0.6ex]
			{LP-SR(2015)} & (10) 0.9758 ± 0.0164 & (9) 37.252 ± 39.7135 & (10) 34.6323 ± 4.6196 & (10) 15 \\[0.6ex]
			{MWGF(2014)} & (8) 0.9800 ± 0.0220 & (5) 24.0500 ± 43.2523 & (6) 36.4995  ± 4.0402 & (7) 24 \\[0.6ex]
			{GFDF(2019)} & (5) 0.9890 ± 0.0134 & (4) 21.4532 ± 34.6367 & (5) 38.5632 ± 5.5568 & (5) 29 \\[0.6ex]
			{MGIMF(2022)} & (4) 0.9929 ± 0.0164 & (13) 8.2240 ± 21.5785 & (3) 42.2132 ± 6.5481 & (4) 35 \\[0.6ex]
			{DRPL(2020)} & (6) 0.9873 ± 0.0119 & (8) 34.6962 ± 39.8551 & (7) 36.2044 ± 7.0005 & (8) 23 \\[0.6ex]
			{IFCNN(2020)} & (7) 0.9854 ± 0.0065 & (2) 17.6740 ± 7.5084 & (8) 36.0513 ± 1.8389 & (6) 25 \\[0.6ex]
			{U2Fusion(2020)} & (13) 0.9012 ± 0.0295 & (6) 253.3413 ± 279.8495 & (14) 25.5515 ± 3.1853 & (14) 4 \\[0.6ex]
			{MFF-GAN(2021)} & (14) 0.8684 ± 0.0613 & (3) 182.1918 ± 181.8785 & (13) 26.4049 ± 2.3061 & (13) 5 \\[0.6ex]
			{DRL-FPD(2022)} & (2) 0.9943 ± 0.0094 & (14) 8.6004 ± 15.7198 & (2) 43.9169 ± 6.9312 & (2) 37 \\[0.6ex]
			{SwinFusion(2023)} & (12) 0.9302 ± 0.023 & (10) 59.2002 ± 22.2597 & (11) 30.7203 ± 1.7039 & (11) 11 \\[0.6ex]
			{MUFusion(2023)} & (11) 0.9485 ± 0.0106 & (12) 76.9215 ± 29.2335 & (12) 29.5937 ± 1.7188 & (12) 10 \\[0.6ex]
			{DB-MFIF(2024)} & (3) 0.9938 ± 0.0051 & (11) 6.8236 ± 5.1374 & (4) 41.0171 ± 3.3967 & (3) 36 \\[0.6ex]
			{Dc-EEMF} & (1) 0.9974  ± 0.0014 & (1) 0.9995  ± 0.3608 & (1) 48.2246 ± 1.0038 & (1) 42 \\[0.6ex]

			\bottomrule[0.3mm]
		\end{tabular}%
	\end{center}
	
\end{table*}%

\begin{table*}[t]
	\setlength{\abovecaptionskip}{0cm} 
	\setlength{\belowcaptionskip}{-1cm}
	\caption{Statistical Metrics (Mean±SD) for Real Phantom and In-vitro Liver Tissue Study}
	
	\label{table4}%
	\begin{center}
		\begin{tabular}{cccccc}
			
			\toprule[0.3mm]
			Metrics & {$Q_E$} & {$Q_{CV}$} & {SD} & {$Q_P$} & {(Rank) Borda count}\\
			\midrule[0.3mm]
			{NSCT-SR(2015)} & (2) 0.8577 ± 0.0658 & (2) 111.7865 ± 73.0474 & (5) 48.2046 ± 16.4331 & (6) 0.5594 ± 0.1354 & (3) 45 \\[0.6ex]
			{LP-SR(2015)} & (3) 0.8543 ± 0.0683 & (3) 113.9816 ± 79.0139 & (4) 48.5376 ± 16.8954 & (7) 0.559 ± 0.1366 & (5) 43 \\[0.6ex]
			{MWGF(2014)} & (4) 0.8452 ± 0.0835 & (5) 162.0033 ± 112.6305 & (2) 49.8418 ± 18.0851 & (1) 0.6739 ± 0.1451 & (2) 48 \\[0.6ex]
			{GFDF(2019)} & (5) 0.8356 ± 0.0850 & (4) 145.6496 ± 118.4772 & (3) 48.7795 ± 17.6876 & (4) 0.5846 ± 0.1395 & (4) 44 \\[0.6ex]
			{MGIMF(2022)} & (6) 0.8299 ± 0.1175 & (8) 291.9402 ± 254.6457 & (6) 48.1892 ± 15.996 & (3) 0.6236 ± 0.1583 & (6) 37 \\[0.6ex]
			{DRPL(2020)} & (12) 0.6857 ± 0.1780 & (13) 665.7343 ± 624.3831 & (8) 44.8876 ± 16.0770 & (12) 0.4454 ± 0.1321 & (12) 15 \\[0.6ex]
			{IFCNN(2020)} & (8) 0.7855 ± 0.0471 & (7) 245.9168 ± 87.5448 & (10) 42.6339 ± 14.7319 & (5) 0.5771 ± 0.1154 & (8) 30 \\[0.6ex]
			{U2Fusion(2020)} & (13) 0.5411 ± 0.0405 & (12) 562.4839 ± 215.9345 & (13) 35.4770 ± 12.6902 & (14) 0.4181 ± 0.1305 & (13) 8 \\[0.6ex]
			{MFF-GAN(2021)} & (11) 0.7346 ± 0.0899 & (10) 367.0852 ± 174.0611 & (9) 44.2794 ± 16.0580 & (9) 0.5567 ± 0.1325 & (9) 21 \\[0.6ex]
			{DRL-FPD(2022)} & (7) 0.8080 ± 0.1115 & (6) 235.2503 ± 237.9049 & (7) 47.0304 ± 15.6026 & (8) 0.5588 ± 0.1254 & (7) 32 \\[0.6ex]
			{SwinFusion(2023)} & (10) 0.7398 ± 0.0681 & (9) 308.8691 ± 166.8041 & (12) 39.0655 ± 13.6472 & (11) 0.4941 ± 0.1504 & (11) 18 \\[0.6ex]
			{MUFusion(2023)} & (14) 0.4772 ± 0.0395 & (14) 854.4542 ± 397.0718 & (14) 31.1913 ± 12.4896 & (13) 0.4236 ± 0.1041 & (14) 5 \\[0.6ex]
			{DB-MFIF(2024)} & (9) 0.7712 ± 0.0806 & (11) 381.8354 ± 238.1086 & (11) 41.8239 ± 13.1402 & (10) 0.5009 ± 0.1129 & (10) 19 \\[0.6ex]
			{Dc-EEMF} & (1) 0.8689 ± 0.0743 & (1) 108.9740 ± 88.2201 & (1) 50.2293 ± 17.8884 & (2) 0.6468 ± 0.1489 & (1) 55 \\[0.6ex]

			\bottomrule[0.3mm]
		\end{tabular}%
	\end{center}
	
\end{table*}%

\begin{figure}[!t]
	\centering
	\includegraphics[width=85mm]{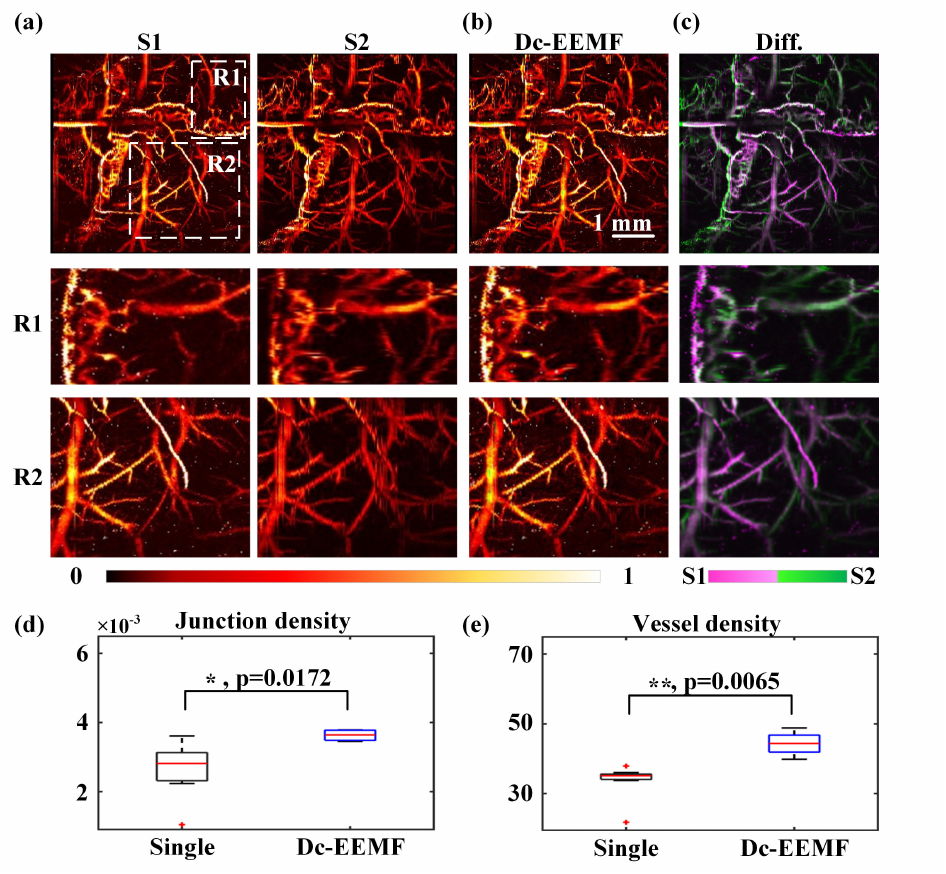}
	\caption{In vivo mouse brain study via Dc-EEMF-powered PAM. (a) and (b) depict paired multi-focus images and the resulting fusion. (c) Represents the difference image between S1 and S2, where pink indicates the region belonging to S1 and green indicates the region belonging to S2. Here, S1 and S2 correspond to source image 1 and source image 2, respectively, while R1 and R2 represent region 1 and region 2. "Diff." denotes the difference image. (d) and (e) display the statistical analysis of junction density and vessel density parameters. \textcolor{black}{Source images are obtained at depths of 0 and 500 µm, respectively.} *, p\textless0.05; **, p\textless0.01; ***, p\textless0.001.}
	\label{figure5}
\end{figure}

\subsection{Real phantom and in-vitro liver tissue study via Dc-EEMF-powered PAM}
\par\textcolor{black}{To demonstrate the preliminary model's capabilities in dealing with real PAM image data, experiments were conducted using the leaf phantom and in-vitro liver tissue. Characterizing the microenvironment related to vasculature plays a crucial role in PAM's medical applications. \textcolor{black}{As our previous model was trained using a mouse brain vasculature dataset, the PAM imaging result of a "vasculature" phantom created with a leaf sample was utilized, displayed in Fig. \ref{figure4}(a) (the first two images on the top row depicting source images 1 and 2).} In the leaf phantom experiment of Fig. \ref{figure4}(a), the information displayed in the single-focus imaging results is limited. 	
The multi-focus image fusion results obtained using the transform-domain-based and end-to-end-based method, including NSCT-SR, LP-SR, IFCNN, U2Fusion, MFF-GAN, SwinFusion, MUFusion, and DB-MFIF, show evident image reconstruction errors in the trunk of the leaf skeleton, indicated by the dashed pink ellipse area. The fusion results obtained with the MGIMF in traditional spatial-domain-based method and DRPL in decision-map-based method exhibit rough details and intermittent structures in the primary leaf skeleton, indicated by the dashed green ellipse area. 
In ROI 1 and ROI 2, the fusion results from the NSCT-SR, LP-SR, MWGF, GFDF, MGIMF, IFCNN, MFFGAN, DRL-FPD, SwinFusion, MUFusion, DB-MFIF methods indicate mis-fusion artifacts as highlighted by the yellow arrows, and the DRPL, DRL-FPD, U2Fusion, MUFusion methods' fusion results indicate missing details as highlighted by the red arrows.
Upon observing the details in the enlarged regions of interest, it becomes evident that the proposed Dc-EEMF method is more proficient in suppressing artifacts in the fused image compared to the others. Additionally, it demonstrates superior information preservation abilities.
The intensity distribution map below the corresponding green dotted line within ROI 2 reveals intensity weakening and intensity mis-fusion in the NSCT-SR, LP-SR, GFDF, DRPL, IFCNN, U2Fusion, MFF-GAN, DRL-FPD, SwinFusion, MUFusion and DB-MFIF methods, the transform-domain-based and end-to-end-based methods, compared with source images, indicate by the yellow arrows.} 
	
\par Furthermore, in Fig. \ref{figure4}(b), experiments on in vitro liver tissues of the hepatic sinusoid demonstrate that the transform-domain-based method could lead to intensity reduction (pink area in IFCNN, U2Fusion, MFF-GAN, MUFusion, and DB-MFIF), while the spatial-domain-based method could lead to the omission of more detailed information (green boxes in MWGF and MGIMF).
From ROI 3, corresponding to each source image, the fusion results in the MGIMF, DRPL, U2Fusion and MUFusion methods exhibit obvious fusion artifacts indicated by yellow arrows and missing information indicated by red arrows.
The intensity distribution map below the corresponding green dotted line within ROI 3 shows intensity degeneration and intensity mis-fusion in the MWGF, IFCNN, U2Fusion, MFF-GAN, DRL-FPD, SwinFusion, MUFusion, DB-MFIF methods compared with S1 or S2. 
The spatial-domain-based method excels in recovering and transferring intensity distribution information from the source images to the fused image, while the transform-domain-based method is better at preventing the degradation of the detailed information in the fused image. The IFCNN method (pink box) tends to exhibit misalignment artifacts, accompanied by noticeable reduction in intensity distribution.
Benefiting from the complementary advantages of these two category methods, the proposed Dc-EEMF method not only preserves the focus intensity distribution of the source image more comprehensively, retaining finer details, but also effectively suppresses various artifacts to a significant extent.

\par\textcolor{black}{For quantitative analysis, in Table \ref{table4}, allowing for the calculation of $Q_E$, $Q_{CV}$, $Q_P$, and SD evaluation metrics. As per the established criteria, quantitative values confirm that Dc-EEMF attains the best fusion performance in terms of $Q_E$, $Q_{CV}$, and SD, with $Q_P$ securing second place. Furthermore, the statistical data distribution from Dc-EEMF maintains higher stability, relatively. Notably, the proposed Dc-EEMF method secures the first place in the Borda count. Hence, the proposed Dc-EEMF exhibits competitive performance compared to other methods, excelling not only in visual perception but also in quantitative analysis.}

\subsection{In vivo mouse cerebral vasculature study via Dc-EEMF-powered PAM}

\par\textcolor{black}{ To further showcase the performance of the proposed Dc-EEMF method, we conducted in vivo imaging of the mouse cerebral vasculature with an intact skull using 6-week-old female BALB/c mice for experiments. Before imaging, the mice were placed under general anesthesia (vaporized isoflurane: $2\%$ for induction and $1.5\%$ for maintenance). The animal's body temperature was maintained at 37$^\circ$C using a fan heater (HD-15, Dyson Supersonic). Subsequently, the mouse's head was secured in a brain stereotactic configuration for further operational preparation. A major portion of the scalp and fascia above the skull was removed to create an imaging window of 5$\times$5 $mm^2$. Before commencing the imaging process, ultrasound gel was applied between the cranial window and the membrane at the bottom of the water tank. Fig. \ref{figure5}(a) showcases different focal points of source images of OR-PAM maps, aiding in the identification of corresponding focus regions. Initially, the system's focal plane was set to the surface of the skull. A two-dimensional raster scanning was executed to capture the depth-resolved photoaocustic signal (source image 1). Subsequently, the focus was shifted downwards by 500 \textmu m, and another two-dimensional raster scanning was conducted to obtain source image 2. The fused image of source images 1 and 2 is depicted in Fig. \ref{figure5}(b). The second and third rows illustrate close-up images of Fig. \ref{figure5}(a), identified by the white rectangles R1 and R2, respectively. Due to the DoF limit of the OR-PAM, only partial details of the vasculature can be fully resolved in each image. However, the fusion models expand the OR-PAM's DoF, enabling a more comprehensive identification of vasculature from various focal points. The fused result obtained using the proposed Dc-EEMF method in Fig. \ref{figure5}(b) displays extensive vasculature information, enhancing the DoF for a clearer visual effect. The difference image, generated using the MATLAB function \textit{imshowpair}, illustrates the variation in focus regions between source image 1 and source image 2 (Fig. \ref{figure5}(c)).} 
\par\textcolor{black}{ In the statistical analysis, the two-tailed Mann-Whitney U Test was employed to quantitatively assess the differences in junction density and vessel density between the compared groups in Fig. \ref{figure5}(d) and (e). Both junction density (U=30, p=0.0172) and vessel density (U=32, p=0.0065) parameters in the fused images surpass those in the single-focused source images. These findings demonstrate the efficacy of the Dc-EEMF-powered method in offering comprehensive characterization of blood vessels, especially in accommodating curvature changes in brain topological structure.}

\begin{figure*}[!b]
	\centering
	\includegraphics[width=150mm]{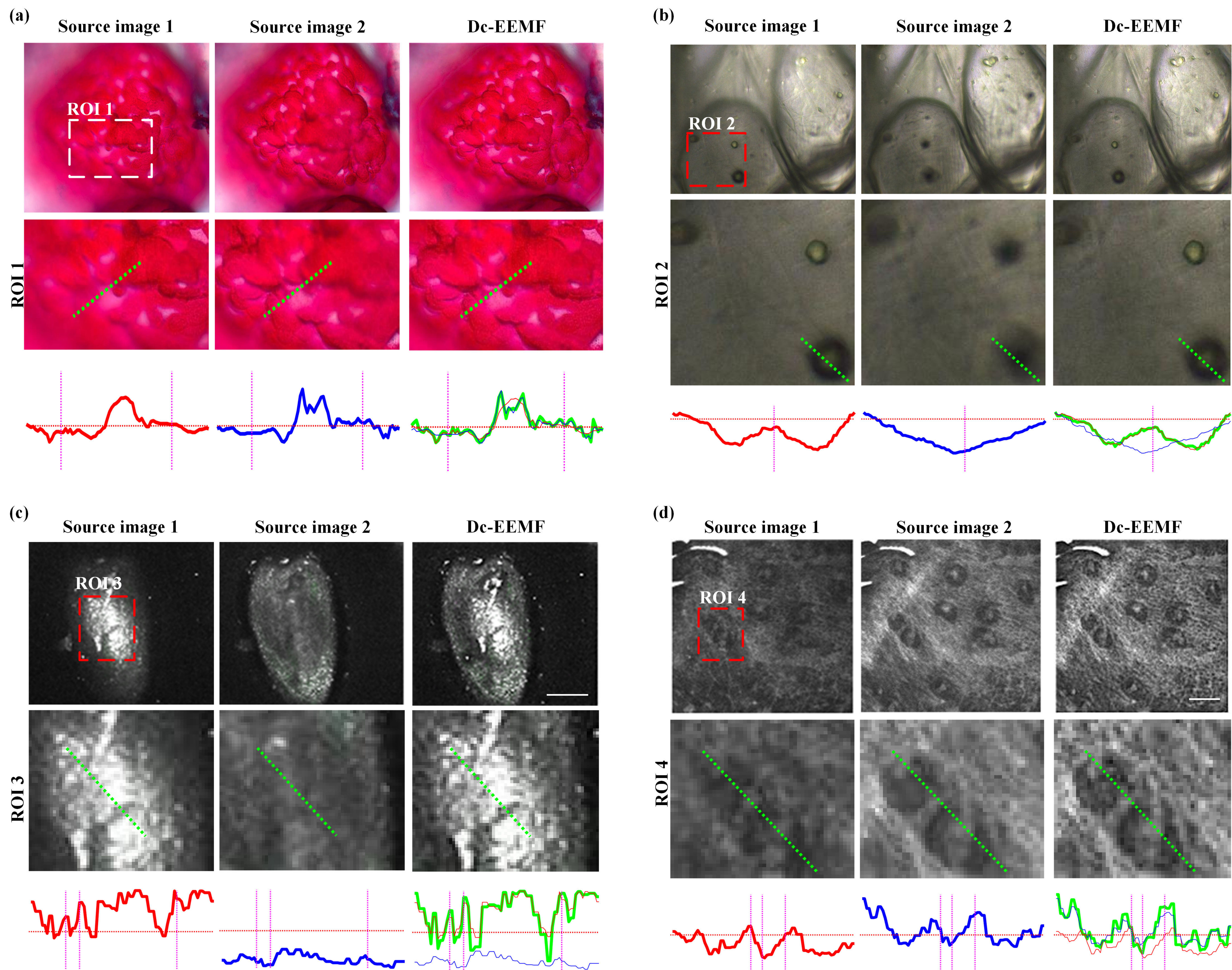}
	\caption{\textcolor{black}{Multi-modalities and applications with proposed Dc-EEMF. (a) A section of mouse intestine containing Peyer's patches, (b)  Live red blood cells in urine sediment sample, (c) Reflectance confocal microscopy imaging of single algal particles (scale-bars correspond to \( 250 \, \mu \)m), (d) Reflectance confocal microscopy imaging of mouse melanoma (scale-bars correspond to \( 140 \, \mu \)m). Source images of (c) are obtained at depths of 150 and 250 µm. Source images of (d) are obtained at depths of 20 and 150 µm.}}
	\label{figure6}
\end{figure*}

\begin{table*}[!t]
	\caption{Statistical Metrics (Mean±SD) for The $L_{dFFP}$ Loss Function Ablation Study}
	\label{table1}%
	\begin{center}
		\begin{tabular}{ccccc}
			\toprule[0.3mm]
			Metrics & {$Q_E$} & {$Q_{CV}$} & {SD} & {$Q_P$} \\
			\midrule[0.3mm]
			{w/o dFPP loss\footnotemark[1]} & (2) 0.8444 ± 0.0868 & (2) 152.6998 ± 129.4994 & (1) 50.4389 ± 15.5137 & (2) 0.8444 ± 0.0868\\[0.6ex]
			{Dc-EEMF} & (1) 0.8526 ± 0.0782 & (1) 141.9502 ± 116.629 & (2) 49.8987 ± 15.5431 & (1) 0.8526 ± 0.0782 \\[0.6ex]
			
			\bottomrule[0.3mm]
		\end{tabular}%

		\footnotemark[1]{w/o dFPP loss represents the Dc-EEMF training only with the $L_{per}$,  $L_{SSIM}$ and ${L}_{FFL}$ three loss functions.}
	\end{center}

\end{table*}%

\begin{table*}[!t]
    \textcolor{black}{
	\caption{Statistical Metrics (Mean±SD) for Feature Fusion Rule Ablation Study}
	}
	\label{table2}%
	\begin{center}
		\begin{tabular}{cccccccc}
			\toprule[0.3mm]
			Metrics & {$Q_E$} & {$Q_{CV}$} & {SD} & {$Q_P$} & {(Rank) Borda count}\\
			\midrule[0.3mm]
			
			{SF} & (5) 0.8505 ± 0.0793 & (2) 147.0997 ± 121.8973 & (4) 49.8727 ± 15.3643 & (5) 0.8505 ± 0.0793 & (4) 8 \\[0.6ex]
			{$c\_w\_max$} & (2) 0.8524 ± 0.0742 & (3) 149.0131 ± 121.3004 & (5) 49.2783 ± 15.3949 & (2) 0.8524 ± 0.0742 & (2) 12 \\[0.6ex]
			{max} & (3) 0.8520 ± 0.0777 & (4) 150.9133 ± 126.8466 & (2) 49.9467 ± 15.4878 & (3) 0.8520 ± 0.0777 & (2) 12 \\[0.6ex]
			{cat} & (4) 0.8520 ± 0.0792 & (5) 154.3041 ± 132.4855 & (1) 50.3466 ± 15.5801 & (4) 0.8520 ± 0.0792 & (3) 10 \\[0.6ex]
			{Dc$-$EEMF} & (1) 0.8526 ± 0.0782 & (1) 141.9502 ± 116.629 & (3) 49.8987 ± 15.5431 & (1) 0.8526 ± 0.0782 & (1) 18 \\[0.6ex]

			\bottomrule[0.3mm]
		\end{tabular}%
		
	\end{center}
\end{table*}%

\subsection{Multi-modalities and pathological application study}
\par \textcolor{black}{To further validate the superiority of the proposed Dc-EEMF method, we conducted additional experiments using optical microscopy and reflectance confocal microscopy datasets \cite{68_microscopydataset}, as shown in Fig. \ref{figure6}.} In Figure \ref{figure6} (a), the Dc-EEMF method accurately preserves the details of the focal region from the optical microscopy source images. In Figure \ref{figure6} (b), it is clearly evident that our method effectively retains the fine details of live red blood cells in a urine sediment sample, especially near the boundary between the focused and defocused regions. In Figure \ref{figure6} (c), the Dc-EEMF keeps a noticeable edge and texture information from the source images, achieving a balanced representation between macro-scale and micro-scale features. In Figure \ref{figure6} (d), due to the high optical absorption properties of melanoma, the complementarity between the two images is reduced. However, the fusion result in ROI 4, obtained by the Dc-EEMF, significantly improves the distinction between the two objects. The proposed Dc-EEMF-powered multi-focus image fusion holds significant value for subsequent analysis. For instance, the accurate and quantitative assessment of vascular parameters from an all-in-focus image, achieved through multi-focus image fusion, is crucial for diagnosing and classifying various vascular diseases, such as hepatic sinusoids, vascular malformations, and brain disorders \cite{40_li_activation_2018,41_moon_intravital_2020}.

\subsection{Ablation studies}
\par\textcolor{black}{ In ablation studies, we use the 20 groups of optimal network parameters in all 120 training epochs. The performances of all models are evaluated by average evaluation metrics on 8 paired of multi-focus photoacoustic microscopy images' dataset in size of 500$\times$500 pixels. Next, we will discuss these ablation studies by qualitative and quantitative comparisons in detail. It can be found that our designed U-Net based loss function has greatly improved the model performance and the channel-wise spatial frequency feature fusion rule makes an impedance against the artifacts.}

\subsubsection{Experimental validation of dFPP loss function}
\par\textcolor{black}{To combine the advantages of the two categories of multi-focus image fusion methods and achieve more accurate fusion results, we designed the dFPP loss function to provide a decision-level constraint information to the end-to-end fusion process. To verify the effectiveness of the dFPP loss function, we compare the fusion results with and without the dFPP loss function. In other words, without the dFPP loss function signifies a purely end-to-end-based method. Our method obtains the best average values for $Q_E$, $Q_{CV}$, and $Q_P$ metrics, as shown in Table \ref{table1}, which demonstrated the superiority of a more accurate fusion results obtained by Dc-EEMF.}

\subsubsection{Experimental validation of channel-wise spatial frequency feature fusion rule}
\par
\begin{figure*}[!t]
	\centering
	\includegraphics[width=150mm]{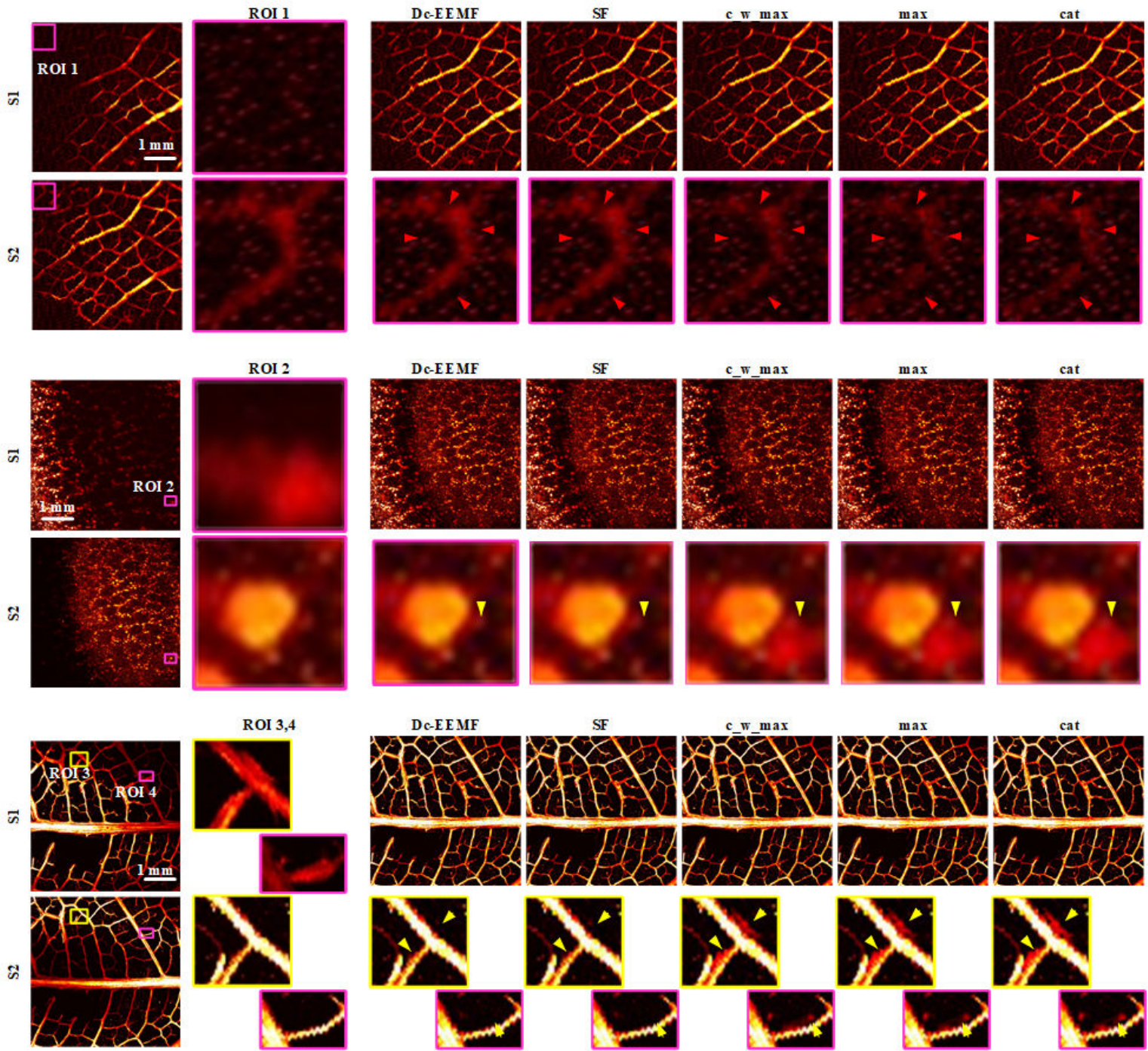}
	\caption{\textcolor{black}{The depth-of-field expansion results for different fusion-rules-based methods. Dc-EEMF employs channel-wise spatial frequency as the feature fusion rule; SF employs the spatial frequency; c$\_$w$\_$max employs the channel-wise window-based maximum; max employs the maximum; cat employs the feature tensors concatenation. Information omission is indicated by red arrows, misalignment artifacts are indicated by yellow arrows. Source images in first two rows obtained at depths of 250 and 750 µm. Source images of middle two rows obtained at depths of 0 and 750 µm. Source images of last two rows obtained at depths of 500 and 750 µm.}}
	\label{figure7}
\end{figure*}

\textcolor{black}{To demonstrate the effectiveness of channel-wise spatial frequency as feature fusion rule for the proposed method, we conducted a comparative analysis involving five different feature fusion rules: the channel-wise spatial frequency (Dc-EEMF), the spatial frequency (SF), the channel-wise window-based maximum (c\_w\_max), the maximum (max) and the feature tensors concatenation (cat). As a result, compared with other methods, Dc-EEMF can better suppress the fusion artifacts caused by mechanical scanning and obtain more accurate fusion results at the trunk of the leaf skeletons, branch detail as well as morphology of hepatic sinuses (Fig. \ref{figure7}). Meanwhile, in the quantitative analysis (Table \ref{table2}), Dc-EEMF obtained the top-two places in $Q_{CV}$, $Q_E$ and $Q_P$. These also prove that by further adding the strategy of channel-wise spatial frequency as the feature fusion rule, the proposed Dc-EEMF method possess more precise fusion results with effectively suppression of the artifact influence. As a result, in Table \ref{table2}, the proposed Dc-EEMF got a Borda count score of 18 points and ranked the first places, which demonstrate the superiority of channel-wise spatial frequency feature fusion-based Dc-EEMF method.}

\section{Discussion and Conclusions}
\par\textcolor{black}{ In this study, we aim to explore the feasibility of computationally addressing pushing DoF limit PAM without introducing additional complexity or cost to the system. To achieve this, we propose a decision-level constrained end-to-end learning method for the multi-focus image fusion task in large-field volumetric imaging of biological tissues with uneven surfaces in OR-PAM. Compared to other multi-focus image fusion algorithms, our Dc-EEMF method introduces a novel approach by combining the complementary advantages of transform domain methods and spatial domain methods. This enables us to achieve a more natural visual effect around boundary regions while preserving the source focus information in specific focused areas, thereby retaining sharpness and contrast (Figs. \ref{figure3} and \ref{figure4}). The superiority of our Dc-EEMF method is both qualitatively and quantitatively evident in the comparisons of simulation, phantom and in vitro tissue results. \textcolor{black}{Furthermore, in vivo mouse brain experiments demonstrated even greater performance, achieving a \( 560 \, \mu\text{m} \) DoF by computationally fusing two focused source images (\( Z = 0 \, \mu\text{m} \quad \text{and} \quad Z = 500 \, \mu\text{m} \)) while preserving acceptable transverse resolution. This represents a substantial improvement over standard OR-PAM systems.} The multi-layer fusion results further validate the superiority of proposed Dc-EEMF method, evidencing high effectiveness and robustness in both visual effects and quantitative metrics( (Fig. \ref{figure6}). Additionally, we conduct an ablation study to verify the effectiveness of our network, specifically examining the U-Net-based loss function strategy and channel-wise spatial frequency feature fusion rule. These enhancements significantly improve the model's performance and mitigate artifacts, respectively. The computed method of Dc-EEMF offers versatile and cost-effective, flexibility, post-acquisition applicability, computational optimization, improved image quality and artifact reduction, as well as reduces the system complexity and eliminates the response time limitations compared with using an electrically tunable lens. From the image acquisition aspect, Dc-EEMF method can achieve fast multi-focus source image acquisition cooperating with MHz laser repetition rates (only two-dimensional multi-focus excitation raster scanning with a span of 2-3 focus layers is required, no axial scanning is required, and no z-stacks are required. \textcolor{black}{The proposed Dc-EEMF method offers significant clinical value by enabling comprehensive quantitative assessment of vascular parameters, including microvascular structure, blood perfusion, oxygenation, and flow dynamics, which is crucial for studying neurological disorders (e.g., stroke, Alzheimer's), characterizing tumor vasculature in oncology, and monitoring atherosclerotic plaque progression. Computationally, the optimized fusion algorithm achieves efficient performance with only 0.24 million parameters, processing \( 128 \times 128 \) size in approximately 29.7 ms on a 2080Ti GPU, representing a 30\% reduction in computational load compared to conventional deep learning-based fusion methods while maintaining superior image quality, making it clinically practical.}}

\par\textcolor{black}{ The proposed Dc-EEMF method demonstrates significant differences in quantitative analysis statistics between leaf skeleton, in vitro liver tissue, and in vivo cerebral vasculature compared to single-focus images $($\cref{figure4,figure5,figure6}$)$. Single-focus images lose some details and cannot provide comprehensive global information. Accurate and quantitative analysis of vascular parameters through multi-focus image fusion proves crucial for comprehensive vascular assessment. It aids in understanding disease mechanisms, distinguishing normal from abnormal vascular structures, quantifying relevant parameters, and monitoring disease progression or treatment response. These analysis techniques, offering objective and quantitative measurements, enhance diagnostic accuracy, aiding better classification and understanding of vascular diseases across various clinical practices. Furthermore, in quantitative analysis of pathological microscopic images requiring pushing DoF limit, multi-focus image fusion significantly improves the accuracy and reliability of subsequent analysis. It extends DoF, integrates information, enhances feature extraction, and ensures consistency. This comprehensive assessment enhances diagnostic accuracy, research outcomes, and understanding of pathological processes. Particularly, applying multi-focus image fusion methods to address tissue surface unevenness in microscopic imaging with high numerical aperture is crucial and pressing. The emergence of optical/photoacoustic microscopy has facilitated artificial intelligence (AI) assisted examination of bone tissue and clinical blood specimens. \textcolor{black}{An extended DoF image is essential for both visual perception and automated AI-assisted analysis, necessitating information fusion from multi-focus layers for comprehensive identification of pathological findings at different depths \cite{42_cao_label-free_2022,43_manescu_content_2022}. From a multi-task learning perspective, we can implement a strategy that utilizes multi-focus cross-modal imaging scanning, combined with multitask multi-focus image fusion and multi-modal image fusion. This integrated approach facilitates both extended DoF and complementary information fusion. The attention mechanisms and feature fusion strategies developed for multi-focus image fusion can be effectively adapted to handle complementary information from various imaging modalities, such as combining pathological slides with functional scans for comprehensive medical analysis.} In the future, integrating this approach with a multispectral PAM system may further aid intravital imaging and diagnosis of vessel-related diseases, quantitative and comprehensive characterization of pathophysiological alterations. This approach provides robust insights into pathophysiological parameters of microvascular structure, blood perfusion, oxygenation, and flow. It holds promise for fundamental studies of cerebral vessel disease, skin cancer, and atherosclerotic plaque, among other diseases.}
\par\textcolor{black}{In conclusion, we have developed a novel Dc-EEMF-powered PAM to push DoF limit via a deep learning method known as decision-level constrained end-to-end learned multi-focus image fusion. A thorough side-by-side comparison conducted in each segment of the validation experiment highlights the advantages of our Dc-EEMF method over conventional computed methods. Notably, it maintains accurate vasculature quantifications across the extended DoF. Ablation experimental results further validate the effectiveness of our proposed method in preserving the original focus information from the source images and managing artifacts around the boundary regions. Looking ahead, we envisage exploring more advanced approaches in the design aspects to enhance the performance of this multi-focus image fusion framework. We believe that this innovative method holds potential for use in large-field volumetric imaging of biological tissues, particularly those with diverse surface topological structures, such as in the study of brain diseases and tumor applications.}

\color{black}



\bibliographystyle{unsrt}
\bibliography{refs}

\end{document}